\documentclass[apl,reprint,superscriptaddress,super,superbib,amsmath,aip]{revtex4-1}
\usepackage{graphicx}
\usepackage{color}
\usepackage{ulem}

\begin{document}  

\title{Accessing the dark exciton spin in deterministic quantum-dot microlenses}

\author{Tobias Heindel}\email{tobias.heindel@tu-berlin.de}
\affiliation{Institut f\"ur Festk\"orperphysik, Technische Universit\"at Berlin, 10623 Berlin, Germany}
\author{Alexander Thoma}
\affiliation{Institut f\"ur Festk\"orperphysik, Technische Universit\"at Berlin, 10623 Berlin, Germany}
\author{Ido Schwartz}
\affiliation{The Physics Department and the Solid State Institute, Technion-Israel Institute of Technology, Haifa 32000, Israel}
\author{Emma R. Schmidgall}
\affiliation{The Physics Department and the Solid State Institute, Technion-Israel Institute of Technology, Haifa 32000, Israel}
\affiliation{Present address: Department of Physics, University of Washington, Seattle WA 98195, USA}
\author{Liron Gantz}
\affiliation{The Physics Department and the Solid State Institute, Technion-Israel Institute of Technology, Haifa 32000, Israel}
\author{Dan Cogan}
\affiliation{The Physics Department and the Solid State Institute, Technion-Israel Institute of Technology, Haifa 32000, Israel}
\author{Max Strau\ss}
\affiliation{Institut f\"ur Festk\"orperphysik, Technische Universit\"at Berlin, 10623 Berlin, Germany}
\author{Peter Schnauber}
\affiliation{Institut f\"ur Festk\"orperphysik, Technische Universit\"at Berlin, 10623 Berlin, Germany}
\author{Manuel Gschrey}
\affiliation{Institut f\"ur Festk\"orperphysik, Technische Universit\"at Berlin, 10623 Berlin, Germany}
\author{Jan-Hindrik Schulze}
\affiliation{Institut f\"ur Festk\"orperphysik, Technische Universit\"at Berlin, 10623 Berlin, Germany}
\author{Andre Strittmatter}
\affiliation{Institut f\"ur Festk\"orperphysik, Technische Universit\"at Berlin, 10623 Berlin, Germany}
\affiliation{Present address: Abteilung f\"ur Halbleiterepitaxie, Otto-von-Guericke Universit\"at, 39106 Magdeburg,Germany}
\author{Sven Rodt}
\affiliation{Institut f\"ur Festk\"orperphysik, Technische Universit\"at Berlin, 10623 Berlin, Germany}
\author{David Gershoni}
\affiliation{The Physics Department and the Solid State Institute, Technion-Israel Institute of Technology, Haifa 32000, Israel}
\author{Stephan Reitzenstein}
\affiliation{Institut f\"ur Festk\"orperphysik, Technische Universit\"at Berlin, 10623 Berlin, Germany}

\begin{abstract}
The dark exciton state in semiconductor quantum dots constitutes a long-lived solid-state qubit which has the potential to play an important role in implementations of solid-state based quantum information architectures. In this work, we exploit deterministically fabricated QD microlenses with enhanced photon extraction, to optically prepare and readout the dark exciton spin and observe its coherent precession. The optical access to the dark exciton is provided via spin-blockaded metastable biexciton states acting as heralding state, which are identified deploying polarization-sensitive spectroscopy as well as time-resolved photon cross-correlation experiments. Our experiments reveal a spin-precession period of the dark exciton of $(0.82\pm0.01)$\,ns corresponding to a fine-structure splitting of $(5.0\pm0.7)\,\mu$eV between its eigenstates $\left|\uparrow\Uparrow\pm\downarrow\Downarrow\right\rangle$. By exploiting microlenses deterministically fabricated above pre-selected QDs, our work demonstrates the possibility to scale up implementations of quantum information processing schemes using the QD-confined dark exciton spin qubit, such as the generation of photonic cluster states or the realization of a solid-state-based quantum memory.\end{abstract}

\maketitle   
The quest for so-called quantum bits (qubits) satisfying the stringent demands of future quantum computation and quantum communication scenarios is actively pursued world-wide \cite{Ladd2010}. In this context, solid-state based matter qubits are of particular interest due to their capability for device integration \cite{Imamoglu1999}. For instance, the coherent properties of bright excitons (BEs) \cite{Bonadeo1998,Zrenner2002} as well as single electron- \cite{Press2008} and hole- \cite{DeGreve2011} spins confined in semiconductor quantum dots (QD) have been extensively explored in recent years. The BE is particular useful as qubit, since its coherent state can be initiated, controlled, and readout \cite{Poem2011} using single picosecond-long optical pulses \cite{Benny2011,Kodriano2012}. The use of the BE for quantum information processing tasks, however, is still limited due to its relatively short radiative lifetime ($\approx$1\,ns). The QD-confined dark exciton (DE), on the other hand, has been demonstrated to constitute an extremely long-lived ($\approx$1$\,\mu$s) \cite{McFarlane2009} matter qubit, which interestingly can also be optically accessed, either in all-opticall \cite{Poem2010} or magneto-optical \cite{Lueker2015} experiments. Similar to the BE, the DE can be initiated \cite{Schwartz2015a} and readout using a short optical pulse, while it features long coherence times ($\approx$100\,ns) \cite{Schwartz2015}. Exploiting these features, the DE was recently used as an entangler for the on-demand generation of entangled multi-photon cluster states \cite{Schwartz2016a}. The optical access to the DE, is enabled via excited biexcitonic states containing two charge carriers with parallel spins \cite{Kodriano2010,Hönig2014}. Such states are usually spin-blockaded from relaxation to the biexciton ground state in which the two electrons and two holes have anti-parallel spins. To date, only one group succeeded in optically accessing the DE spin via biexcitonic spin-triplet states \cite{Poem2010}. This first demonstration used a simple planar and non-deterministic sample, offering only limited photon extraction. To push experiments beyond the proof-of-principle stage, however, requires larger photon harvesting efficiencies only achievable with advanced photonic microstructures. Additionally, the sample material used in Ref. \cite{Poem2010} and subsequent work was grown by molecular beam epitaxy (MBE) exclusively. A proof, that the scheme of optically accessing the DE spin is possible also in devices grown with other growth techniques is still pending. For these reasons, scalable implementations of the scheme presented in Ref. \cite{Poem2010} remained elusive so far.

In this work, we employ deterministically fabricated microlenses with integrated QDs grown by MOCVD to optically prepare and readout the DE spin qubit and observe its coherent precession. The use of monolithic microlenses is highly beneficial in this type of experiment, as the lens provides broadband enhancement of photon extraction for the required spectrally separated optical transitions. Our experiments reveal the precession of the DE's spin, from which we are able to deduce the fine-structure splitting between the DE eigenstates. Achieved with deterministic photonic microstructures based on MOCVD-grown material, the results presented in the following demonstrate the robustness of the scheme of optically accessing the DE spin qubit against specific growth conditions and device fabrication methods. In particular, they provide promises for scalable implementations of quantum information schemes using the QD-confined DE as solid-state-based spin qubit.

The samples utilized for our experiments were grown by MOCVD on a GaAs (001) substrate. A layer of self-organized InGaAs QDs, located above a distributed Bragg reflector with 23 pairs of AlGaAs/GaAs layers, is capped with 400\,nm of GaAs. The capping layer provides the necessary semiconductor material for the microlens fabrication via 3D in-situ electron-beam lithography [17], where preselected target QDs are deterministically integrated within monolithic microlens structures (see illustration in Fig. 1(a)). For details on the sample layout and the microlens fabrication, we refer to Refs. \cite{Gschrey2015b} and \cite{Gschrey2013}. The use of a geometric lens approach, providing efficient photon extraction from our target QD in a wide spectral range ($\approx$20\,nm), is a great advantage compared to e.g. cavity approaches exploiting narrow-band enhancement, for devices requiring simultaneous access to multiple, energetically separated states. Optical experiments were performed in a confocal micro-photoluminescence ($\mu$PL) setup with the sample mounted onto the cold finger of a cryostat at temperatures in the range of 4\,K to 20\,K. The sample is excited using a wavelength-tunable continuous-wave Titanium-sapphire laser at a wavelength between 850 and 880 nm, corresponding to wetting-layer excitation for the present QDs. The PL from the QD microlenses is collected via a microscope objective (numerical aperture (NA): 0.40 or 0.85) and spectrally analyzed using grating spectrometers (spectral resolution: 25\,$\mu$eV). Correlations between photons emitted from different spectral lines are studied by means of polarization sensitive intensity cross-correlation measurements \cite{Regelman2001,Kodriano2014}, where polarization optics allow for the detection of photons in any desired polarization basis. The light at the monochromators' exit slits is detected either by two fiber-coupled Silicon-based single-photon counting modules (SPCMs) or by two nanowire-based superconducting single-photon detectors (SSPDs) with a timing resolution of 250\,ps and 90\,ps, respectively. Time resolved coincidence measurements are enabled using time-correlated single-photon counting electronics. 

Figure 1(a) shows the $\mu$PL spectra of a deterministic QD microlens recorded in rectilinear (HV) polarization bases at a temperature of $T=7\,$K.
\begin{figure}[htbp]
\includegraphics[width=\linewidth]{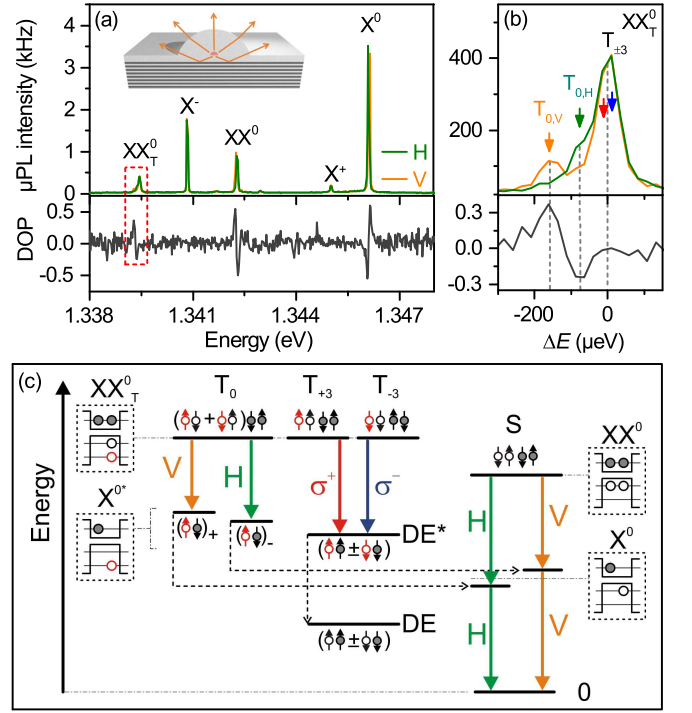}
\caption{\label{Fig_1} (a) Polarization sensitive micro-photoluminescence ($\mu$PL) spectra (upper panel) and extracted rectilinear (HV) degree of polarization (DOP) (lower panel) of a single QD microlens. The relevant excitonic states are labeled: bright exciton (X$^0$), biexciton (XX$^0$), charged trions (X$^-$ and X$^+$) and emission of the spin-blockaded biexciton triplet (XX$_{\rm{T}}^0$). Inset: Schematic of a deterministic microlens used for our experiments. A single QD is integrated within a monolithic microlens above a lower Bragg mirror. (b) PL from the spin-blockaded biexciton triplet states (XX$_{\rm{T}}^0$) on an expanded energy scale, revealing two cross rectilinearly polarized components resulting from the XX$_{\rm{T}}^0$ biexciton and twofold degenerate unpolarized XX$_{\rm{T}\pm3}^0$ biexcitons. (c) Schematic description of the biexcitons´ and excitons´ energy levels and optical transitions illustrating the selection rules observed in (a) and (b). Empty (filled) circles indicate holes (electrons) and solid (dashed) arrows represent radiative (non-radiative) relaxation processes. Red (black) empty circles indicate holes located in the QD's p-shell (s-shell).}
\end{figure}
Various spectral lines are observed and identified using their typical polarization- and excitation intensity dependencies. Emission of the bright exciton (X$^0$), the ground-state biexciton (XX$^0$) as well as the negatively and the positively charged trions (X$^-$, X$^+$) are identified. The polarization selection rules of these spectral lines can be inferred from the lower panel of Fig. 1(a), where the degree of polarization (DOP) in the rectilinear basis is presented according to $\rm{DOP}=(I_{\rm{H}}-I_{\rm{V}})/(I_{\rm{H}}+I_{\rm{V}})$, where $I_{\rm{H}}$ and $I_{\rm{V}}$ refer to the $\mu$PL intensity in H and V polarization, respectively. Both, the X$^0$ and XX$^0$ lines are composed of two cross-linearly-polarized components, exhibiting a fine-structure splitting of $\Delta E_{\rm{FSS,BE}}=(36\pm1)\,\mu$eV. The trion lines X$^-$ and X$^+$, on the other hand, are unpolarized, as expected. At the low energy tail of the spectrum additional spectral lines are visible at 1.3395\,eV. These lines are more clearly seen in the expanded energy scale of Figure 1(b), which reveals a spectral triplet composed of one unpolarized line and two, about a factor of 4 weaker, cross-rectilinearly polarized lines. As explained in the following, these excitonic features can be attributed to emission from the biexcitonic spin-triplet states XX$_{\rm{T}}^0$ (suberscript indicating the respective spin configuration). In contrast to the common spin-singlet biexciton state XX$^0$, the biexcitonic triplet states are constituted of two s-shell electrons and two holes, whereof one hole is in the s-shell while the other one is in the p-shell. The possible spin-configurations of the triplet and singlet states are illustrated in the energy level scheme in Fig. 1(c). In case of antiparallel hole spins, the total spin vanishes and the respective state XX$_{\rm{T}0}^0$ radiatively decays via recombination of a s-shell electron-hole pair under the emission of one H- or V-polarized photon. Subsequently, the QD is left in either the symmetric or antisymmetric superposition of the excited exciton states X$_{+/-}^{0*}$, which reveal a fine structure splitting similar to the bright exciton states X$^0$. The X$_{+/-}^{0*}$ states quickly relax non-radiatively to the bright exciton states X$^0$, which in turn decays radiatively by emitting one V- or H-polarized photon. Therefore, the biexcitonic triplet state XX$_{\rm{T}0}^0$ the bright exciton state X$^0$ and the ground state (empty QD) constitute a radiative cascade, similar to the XX$^0$-X$^0$ cascade. Importantly, the two photons emitted by the XX$_{\rm{T}0}^0$-X$^0$ cascade are cross-rectilinearly correlated with respect to their polarization, due to the underlying selection rules \cite{Kodriano2010}. This behavior is opposite to the common XX$^0$-X$^0$ cascade, which is co-rectilinearly correlated. This picture changes if the initial spin-triplet state is constituted of two holes with parallel spin-projection. The two resulting states XX$_{\rm{T}\pm3}^0$ are almost degenerate in energy and have a total spin projection of ±3 on the QD growth axis. In this case, the radiative recombination of the s-shell electron-hole pair results in the emission of a left- (L) or right- (R) hand circularly polarized photon, leaving the QD in an excited dark exciton state (DE*) with parallel spin-configuration. After a fast spin-preserving non-radiative relaxation of the hole from the p- to the s-shell, the QD ends in the DE ground state X$_{\rm{DE}}^0$. Therefore, the detection of a photon stemming from the XX$_{\rm{T}\pm3}^0$-DE* transition heralds the formation of the DE state \cite{Poem2010}.

To verify the selection rules and correlations described above, we performed polarization-sensitive time-resolved photon correlation measurements on both biexciton cascades using the SPCM detectors. The excitation power is hereby set to a point where the PL from the XX$^0$ line is approximately half of that from the X$^0$ line ($P=9\,\mu$W). Figure 2(a) shows the cross-correlations of the biexciton-exciton XX$^0$-X$^0$ emission as a function of the time between detection events, where the biexciton photon detection is chosen to trigger the coincidence counting experiment. 
\begin{figure}[htbp]
\includegraphics[width=\linewidth]{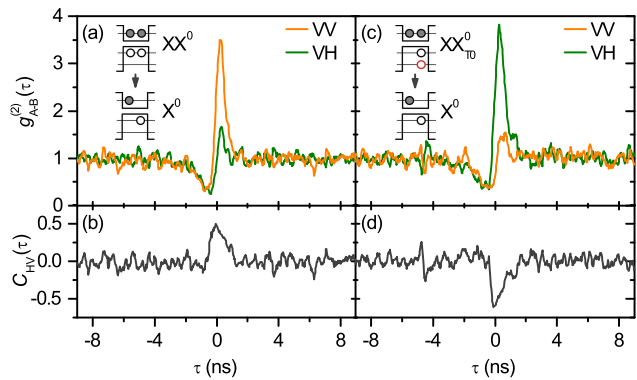}
\caption{\label{Fig_2} Polarization sensitive measurements of the photon cross-correlation $g^{(2)}_{\rm{{A-B}}}(\tau)$ and extracted degree of rectilinear (HV) polarization correlation $C_{\rm{HV}}(\tau)$ for the two different radiative cascades: (a) and (b) the spin-singlet biexciton-exciton (XX$^0$-X$^0$) cascade and (c) and (d) the spin-triplet biexciton-exciton (XX$_{\rm{T}0}^0$-X$^0$) cascade. Schematics show the charge-carrier configuration of the QD initial states leading to the respective photon detection. Strong bunching is observed in both cases confirming the biexcitonic origin of the initial states. For the XX$_{\rm{T}0}^0$-X$^0$ cascade bunching is observed in cross-rectilinear polarization measurements (VH), in contrast to the XX$^0$-X$^0$ cascade, confirming the selection rules and energy level alignment illustrated in Fig. 1(c).}
\end{figure}
The radiative cascade reveals itself in a pronounced bunching in the co-rectilinearly polarized coincidences (VV), while the cross-rectilinearly polarized coincidences (VH) are strongly suppressed. This results in a positive degree of polarization correlation $C_{\rm{HV}}(\tau)=(g^{(2)}_{\rm{VV}}-g^{(2)}_{\rm{VH}})/(g^{(2)}_{\rm{VV}}+g^{(2)}_{\rm{VH}})$ during the radiative cascade (cf. Figure 2(b)). Similar behavior is observed in Figure 2(c), where photon coincidences resulting from the indirect radiative cascade of XX$_{\rm{T}0}^0$ and X$^0$ are presented. Again a pronounced bunching signifies the biexcitonic origin of the initial state XX$_{\rm{T}0}^0$. The polarization selection rules, however, are reversed in this case. As a result, a negative degree of polarization correlation ($C_{\rm{HV}}<0$) is obtained in Figure 2(d). This behavior confirms the selection rules illustrated in Figure 1(c) in agreement with the observations reported in Ref. \cite{Kodriano2010}, which allows us in the following to optically access the DE via the spin-blockaded biexciton state with $\pm3$ spin projection.

Next, we demonstrate that we can all-optically prepare and readout the DE's spin state. For these experiments we employed SSPDs, enabling an improved timing resolution (cf. setup description). The preparation of the DE is performed by photon detection of the XX$_{\rm{T}\pm3}^0$ state, which heralds the presence of the DE state. The readout is performed by adding a single charge carrier, i.e. electron or hole, to the QD. This converts the DE to an optically bright trion state, X$^-$ or X$^+$, which recombines radiatively leaving a single charge carrier in the QD. In our experiment, we add single electrons or holes via  spontaneous charging, and hence cross-correlate the XX$_{\rm{T}\pm3}^0$ emission with photons from the trion state X$^-$ or X$^+$, respectively. Figure 3(a) presents the measured cross-correlation between the XX$_{\rm{T}\pm3}^0$ and the X$^+$ trion for co- and cross-circular polarizations.
\begin{figure}[htbp]
\includegraphics[width=\linewidth]{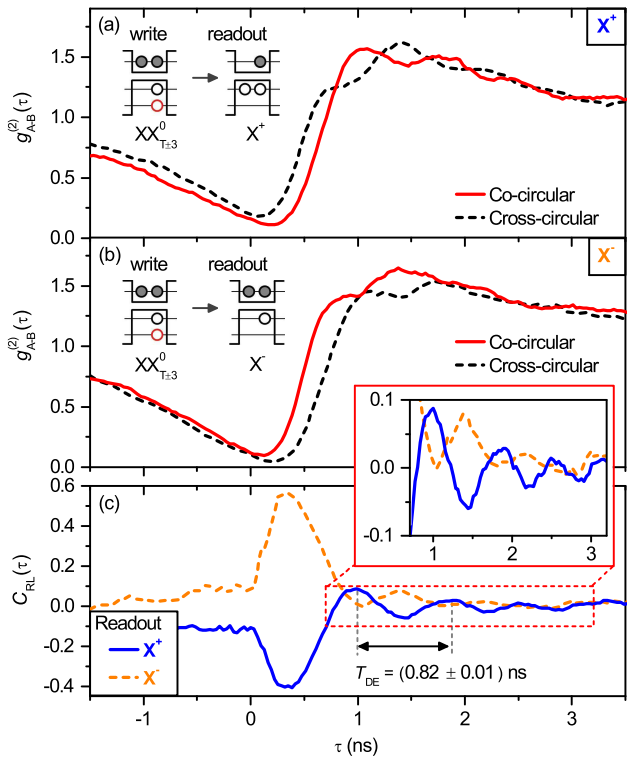}
\caption{\label{Fig_3} Photon cross-correlation experiments between the XX$_{\rm{T}\pm3}^0$ state (dark exciton (DE) preparation) and the charged trion states (DE read-out), (a) X$^+$ and (b) X$^-$ (cf. schematics for the respective charge-carrier configuration). The solid and dashed line corresponds to co- (LL+RR) and cross- (RL+LR) circular polarization, respectively. (c) Degree of correlation in circular polarization $C_{\rm{RL}}(\tau)$ extracted from the measurement data presented in (a) and (b). Due to the precession of the dark exciton (DE) spin between preparation and readout, oscillations occur with a period $T_{\rm{DE}}$ revealing a DE fine-structure splitting of $(5.0\pm0.7)\,\mu$eV.}
\end{figure}
The experimental conditions, i.e. $P=2.0\,\mu$W, $\lambda=877.2\,$nm and $T=17.7\,$K, were carefully chosen to achieve comparable detection rates from both spectral lines while simultaneously providing low population of the bright exciton. We observe a pronounced photon bunching at positive delay times, resulting from the QD charging, which converts the DE into the bright trion X$^+$. The correlated signal at positive times exhibits oscillations with opposite phase for co- and cross-circular polarizations. These oscillations result from the coherent precession of the DE's spin as reported in Ref. \cite{Poem2010}. Due to a finite fine-structure splitting $\Delta E_{\rm{FSS,DE}}$ between the DE eigenstates $\left|\uparrow\Uparrow\pm\downarrow\Downarrow\right\rangle$ \cite{Bayer2002b}, a coherent superposition of these states, heralded by the detection of a circularly polarized XX$_{\rm{T}\pm3}^0$ photon, precesses in time with a period of $T_{\rm{DE}}=h/E_{\rm{FSS,DE}}$, where $h$ is Planck's constant. Similar observation is obtained for the case in which the readout of the DE is performed via the negatively charged trion X$^-$, as presented in Figure 3(b) ($P=1.8\,\mu$W, $\lambda=877.2\,$nm and $T=9.5\,$K). The phase of the oscillations for co- and cross-circular polarization correlations, however, is reversed compared to the readout via the X$^+$ state. This becomes more evident in Figure 3(c), where the degree of circular correlation $C_{\rm{RL}}(\tau)$ deduced from the experimental data shown in (a) and (b) is displayed. Here we are able to observe up to 4 complete spin-precession cycles of the DE state. Fitting the experimental data with an exponentially damped cosine function, reveals a precession period of the DE of $T_{\rm{DE}}=(0.82\pm0.01)\,$ns. This corresponds to a fine-structure splitting of  $\Delta E_{\rm{FSS,BE}}=(5.0\pm0.7)\,\mu$eV. The relatively short precession time of the DE observed in our experiment, would in principle allow for an about 4-times faster rate of entangled-photon generation compared to that reported in Ref. \cite{Schwartz2016a} where the DE has been used as an entangler with an intrinsic period of $T_{\rm{DE}}\approx3\,$ns. 
The demonstrated all-optical preparation and readout of the DE spin using deterministic QD microlenses is an important step towards advanced quantum information schemes exploiting the DE spin-qubit. While previous experiments were carried out with non-deterministic MBE-grown planar devices suffering from limited photon-extraction efficiencies, our work proves the scalability of implementations exploiting the QD-confined DE spin-qubit. Moreover, by clearly observing quantum beats of the DE spin in MOCVD-grown QDs, we provide evidence for the robustness of the DE's spin properties against specific growth conditions, and therefore answer the so far open question, whether this approach is a potentially fragile effect difficult to implement. Employing further technological improvements, such as photonic microstructures with a backside gold mirror \cite{Fischbach2017}, we expect photon extraction efficiencies of up to 80\% for large-NA collection optics, e.g. enabled by on-chip integration of micro-objectives \cite{Fischbach2017a}.

In summary, we exploited a deterministically fabricated QD microlens to optically access the DE spin state. The spin-state is heralded by detecting photons from the spin-blockaded biexciton, while the readout is performed by detecting photons emitted by either the positively or negatively charged trions. In both cases, we clearly observe the coherent precession of the DE spin with a period of $(0.82\pm0.01)\,$ns corresponding to a fine-structure splitting of $(5.0\pm0.7)\,\mu$eV. By employing deterministic microlenses defined above pre-selected QDs, our work shows promise for the transfer of first proof-of-principle experiments on photonic cluster state generation \cite{Schwartz2016a} towards scalable implementations. Exploiting the DE spin qubit in deterministic QD microlenses supplied with electrical contacts \cite{Schlehahn2016a}, could lead to significantly improved entanglement fidelities compared to protocols based on all-optical pulse sequences \cite{Schmidgall2015}.\\
\\
\textbf{Funding:} This work was financially supported by the German Science Foundation within the Collaborative Research Center CRC 787 'Semiconductor Nanophotonics: Materials, Models, Devices' and the German-Israeli-Foundation for Scientific Research and Development (GIF), Grant-No.: 1148-77.14/2011. T.H. acknowledges support by the COST Action MP1403 'Nanoscale Quantum Optics' via a STSM-Grant.\\
\\
\textbf{Acknowledgement:}
Expert sample preparation by R. Schmidt and technical support by C. Hopfmann is gratefully acknowledged.


%

\end{document}